\documentclass[prd,aps,floats,preprintnumbers,preprint]{revtex4}

\usepackage{graphicx}
 
\newcommand{\be}{\begin{equation}}
\newcommand{\ee}{\end{equation}}
\newcommand{\bea}{\begin{eqnarray}}
\newcommand{\eea}{\end{eqnarray}}

\begin{document}


\setlength{\unitlength}{1mm}

\title{What could the value of the cosmological constant tell us about the future variation of the fine structure constant?
\begin{center}



\end{center}
}
\author{Antonio Enea Romano$^{1,2}$}
\affiliation{
${}^{1}$Yukawa Institute for Theoretical Physics, Kyoto University, Kyoto 606-8502, Japan;\\
${}^{2}$Department of Physics, McGill University, Montr\'eal, QC H3A 2T8, Canada \\
}



\begin{abstract}
Motivated by reported claims of the measurements of a variation of the fine structure constant $\alpha$ we consider a theory where the electric charge, and consequently $\alpha$, is not a constant but depends on the Ricci scalar $R$. 
We then study the cosmological implications of this theory, considering in particular the effects of dark energy and of a cosmological constant on the evolution of $\alpha$. 
Some low-red shift expressions for the variation of $\alpha(z)$ are derived, showing the effects of the equation of state of dark energy on $\alpha$ and observing how future measurements of the variation of the fine structure constant could be used to determine indirectly the equation of state of dark energy and test this theory. In the case of a $\Lambda CDM$ Universe, according to the current estimations of the cosmological parameters, the present value of the Ricci scalar is $\approx 10\%$ smaller than its future asymptotic value determined by the value of the cosmological constant, setting also a bound on the future asymptotic value of $\alpha$. 
\end{abstract}

\maketitle

There have been different claims about the detection of the variation \cite{Murphy:2003hw,King:2012id,Murphy:2006vs,Tzanavaris:2006uf,Uzan:2010pm} of the fine structure constant.
Different models have been proposed to account for such a phenomenon, in particular theories where the electric charge is treated as free field \cite{Bekenstein:1982eu}.  We consider the cosmological implications of a model in which the electric charge depends on the local value of the space-time curvature..
This model can be considered a special case of the Bekenstein theory \cite{Bekenstein:1982eu}, where the electric charge depends on the  Ricci scalar, and consequently it is not necessary to introduce an additional kinetic term in the Lagrangian to determine the dynamics of the field associated to the  electric charge, as in the standard Bekenstein case.
In this sense this model has the advantage of not introducing a new fully independent degree of freedom and can consequently give more definite predictions.
At the same time the hypothesis that the electric charge depends only on the Ricci scalar is in agreement with the principle of general covariance and as such could be extended to the study of the variation of other natural constants without the need of introducing any new degree of freedom.


On the basis of general covariance it is natural to consider as an extension of the standard electromagnetism an action of the type 

\bea
S&=&\int{\sqrt{-g}(f(R)F_{\mu\nu}F^{\mu\nu}} \label{fRF} + e_0 u_{\mu}A^{\mu})\,,
\eea
where $F_{\mu\nu}=\partial_{\mu}A_{\nu}-\partial_{\nu}A_{\mu}$ is the Faraday tensor and $A{\mu}$ is the vector potential, as defined in the standard covariant formulation of electromagnetism.

The main difference respect to the classical formulation of electromagnetism corresponds to the fact that such an action implies a space time variation of the fine structure constant $\alpha$.
This can be seen easily by varying the action respect to the vector potential $A^{\mu}$, and after neglecting derivative terms of quantities varying only on very large scales we get
\bea
\partial_{\mu}F^{\mu\nu}&=&f(R)^{-1}e_0 u^{\nu}\,,   
\eea  
which shows how the effective electric charge appearing on the r.h.s. is not constant.
The variation of the action respect to the metric is not expected to modify the standard Einstein's equations because the term $F_{\mu\nu}F^{\mu\nu}\propto (E^2-B^2)$ should give an average vanishing contribution in the case of a homogenous electromagnetic field.
The above action can be considered a particular case of the Bekenstein theory :

\bea
S_{B}&=&\int{\sqrt{-g}(\epsilon^{-2}F_{\mu\nu}F^{\mu\nu}}+e(x_{\mu})  \, u_{\mu}\tilde{A}^{\mu} )=\int{\sqrt{-g}(\epsilon^{-2}F_{\mu\nu}F^{\mu\nu}}+e_0  u_{\mu}A^{\mu} ) \,,\\
A_{\mu}&=&\epsilon \tilde{A}_{\mu} \,,\\
F_{\mu\nu}&=& \partial_{\mu} A_{\nu}-\partial_{\nu}A_{\mu} \,,
\eea
where the electric charge appearing in the Lagrangian is treated as a free field 
\bea
e(x_{\mu})&=&e_0 \epsilon(x^{\mu})\,,
\eea
and the action (\ref{fRF}) corresponds to the case in which 
\bea
\epsilon(x^{\mu})&=&{f(R)}^{-1/2} \,, \\
\alpha(z)&=&\alpha_0 f(R)^{-1}\,.
\eea
In the Bekenstein theory an additional kinetic term is introduced in the Lagrangian in order to determine the dynamics of $\epsilon$, while for the action S (\ref{fRF}) this is not necessary since $f(R)$ entirely determines the variation
of the fine structure constant.
From a physical point of view this corresponds to assuming that the fine structure constant value depends only on the Ricci scalar, preserving the invariance under general coordinate transformations.


The Ricci scalar can be related to the trace of the energy momentum tensor by taking the trace of the Einstein tensor:
\bea
R&=&-T_{\mu}^{\mu}\,,
\eea
For a set of perfect fluids with equation of state
\bea
P_i&=&w_i\rho_i\,,
\eea
we get
\bea
R=\sum_i(3w_i-1)\rho_i \label{Ricci}\,,
\eea
where the index i stands for the i-th component of the total energy density of the Universe. 
It is interesting to note that, according to the above equation, radiation does not contribute to the Ricci scalar at any time, since it vanishes for $w_i=1/3$.

We can now derive a low red-shift expansion of the variation of the fine structure constant.
Even if this would not be valid at higher re-shift, it can still give some useful insight about the features of the theory we are studying.
 In order to get the leading corrections to the value of $\alpha$ we expand the equation of state of dark energy $w(z)$ and the function $f(R)$ according to:
\bea
\Delta\alpha&=&\frac{\alpha(z)-\alpha_0}{\alpha_0}=\Delta\alpha_1 z+ \Delta\alpha_2 z^2+ ..\,, \\
w(z)&=&w_0+w_1 z+w_2 z^2 + .. \,,\\
f(R)&=&1+f_1\left(\frac{R-R_0}{R_0}\right)+f_2 \left(\frac{R-R_0}{R_0}\right)^2+ .. \,,
\eea
where
\bea
R_0&=&R(z=0)\,.
\eea
is the present time value of the Ricci scalar \,,
The above expansion for $f(R)$ is clearly dimensionally consistent and satisfy the normalization condition 
\bea
\alpha(z=0)=\alpha_0\,.
\eea
It should also be noted that such a local expansion is only valid at relatively recent times otherwise, at early times, when $(R-R_0)/{R_0}\rightarrow\infty$, $\alpha$ would diverge.
Nevertheless it is a good approximation to estimate the future variation of $\alpha$ because $(R_{\infty}-R_0)/{R_0}\approx -0.1$, as it will be shown later.

Assuming a flat Universe, after substituting in eq.(\ref{Ricci}) the contributions from matter and dark energy 
\bea
\rho_X(z)&=&3H_0^2 \Omega _X \exp\Bigg[\int_0^z \frac{3 [1+w(x)]}{1+x} \, dx\Bigg] \,,\\
\rho_M(z)&=&3H_0^2(1+z)^3\,,
\eea
we get the following expression for the Ricci scalar as a function of the red-shift
\bea
R(z)&=&-3 (1+z)^3 H_0^2 \Omega _M+3H_0^2 \Omega _X \exp\Bigg[{\int_0^z \frac{3 [1+w(x)]}{1+x} \, dx}\Bigg]  [3 w(z)-1]\,.
\eea
The above expression is also valid at early times when the Universe was radiation dominated since, as noted previously, radiation does not contribute at any time to the Ricci scalar.

We can then compute the leading corrections to the value of the fine structure constant as a function of the red-shift:
\bea
\Delta\alpha_1&=&\frac{3 f_1 (-1+2 w_0 \Omega _X+3 w_0^2 \Omega _X+w_1 \Omega _X)}{3 w_0 \Omega _X-1}\,,\\
\Delta\alpha_2&=&\frac{1}{{2 (1-3 w_0 \Omega _X){}^2}}\Bigg[3 \Big((6 f_1^2-3f_2) (-1+2 w_0 \Omega _X+3 w_0^2 \Omega _X +w_1 \Omega _X){}^2 + \,,\\
&&-f_1 (1-3 w_0 \Omega _X) (2-(-2+24 w_0^2+18 w_0^3+4 w_1+2 w_0 (1+6 w_1)+w_2) \Omega _X)\Big)\Bigg]\,, \nonumber  
\eea
where we have used the relation 
\bea
\Omega_{\Lambda}+\Omega_X=1\,,
\eea
following from the Friedman equation evaluated at present time.
As it can be seen the linear contribution to the variation of the fine structure constant value is affected by the central value and first order derivative of the equation of state of dark energy and by the first derivative of function $f(R)$.
Conversely the above expression shows how measurements of the variation of the fine structure constant could be used to determine indirectly the equation of state of dark energy.

Another important case to consider is that of a $\Lambda CDM$ Universe,  for which the Ricci scalar ass a function of the red-shift is given by
\bea
R(z)&=&-\rho_M^0(1+z)^3-4\rho_{\Lambda}=-3H_0^2(\Omega^0_M(1+z)^3+4\Omega_{\Lambda})\,,
\eea
which implies that at late times $R(z)$ tends to a constant value determined by the cosmological constant, while at earlier times it decreases.

We can then obtain a local expansion for $\alpha(z)$ as in the dark energy case:

\bea
\Delta\alpha_1&=&\frac{3 f_1 (-1+\Omega _{\Lambda })}{1+3 \Omega _{\Lambda }} \,,\\
\Delta\alpha_2&=&\frac{3 (-1+\Omega _{\Lambda }) (6 f_1^2 (-1+\Omega _{\Lambda })-3 f_2 (-1+\Omega _{\Lambda })+f_1 (2+6 \Omega _{\Lambda }))}{2 (1+3 \Omega _{\Lambda }){}^2} \,.
\eea
It can be easily checked that these last formulae are in agreement with the dark energy case when $\{w_n=0,w_0=-1\}$.
In this case the variation of $\alpha$ is completely determined by the coefficients $\{f_1,f_2\}$ and the value of the cosmological constant.

Given the current estimation of cosmological parameters \cite{Ade:2013lta,Ade:2013xsa,Planck:2013qta} the Ricci scalar is tending to an  asymptotic value determined by the cosmological constant
\bea
\frac{R_{\infty}}{R_0}&=&\frac{4 \Omega _{\Lambda }}{1+3 \Omega _{\Lambda }}\approx 0.9 \,,
\eea
and the present value is quite closed to the asymptotic one as it can be seen in Fig. 1.
\begin{figure}
	\centering
		\includegraphics[height=60mm,width=80mm]{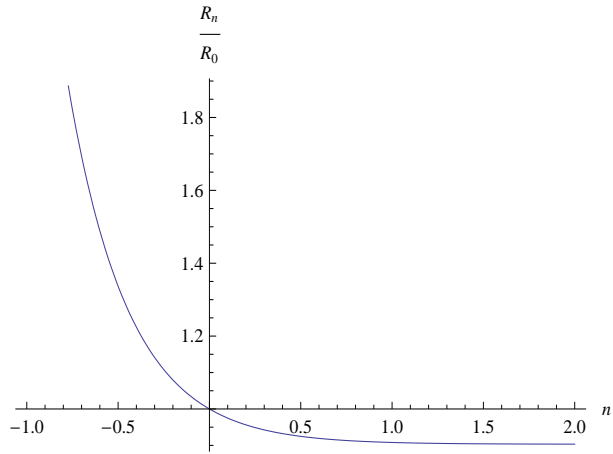}
		\caption{ 
		The ratio $R_n/R_0$ is plotted as a function of the number $n$ of e-folds from today, where $R_n=-3H_0^2(\Omega^0_M(	  \frac{a_0}{a})^3+4\Omega_{\Lambda})$, $a=a_0 e^n$ and $R_0$ is the value of the Ricci scalar today, when $n=0$.
		As it can be seen the present value $R_0$ is quite closed to the asymptotic future value of R.}
	\label{fig:Rn}
\end{figure}
This shows how the Ricci scalar is not expected to vary largely in the future history of the Universe, and if the theory we are studying is correct, also the fine structure constant will reach a corresponding asymptotic value given by
\bea
\alpha_{\infty}&\approx &\alpha_0f(R_{\infty})^{-1}\approx \alpha_0[1+f_1(-0.1)+f_2(-0.1)^2]^{-1}\approx \alpha_0[1+f_1(0.1)-f_2(0.1)^2] \,.
\eea

According to this prediction the so called coincidence problem, i.e. the fact that we happen to live just around the time of transition between a matter dominated and a cosmological constant dominated stage of the evolution of the Universe, would  also have implications on the future variation of the fine structure constant, which would be expected to tend to an asymptotic value not too different from the present one.
After the coefficients ${f_1,f_2}$ have been determined by fitting available observational data of $\Delta\alpha(z)$ we could get a good estimation of $\alpha_{\infty}$, but we would have to wait a time interval of the order of the Hubble time $H_0^{-1}$ before the matter contribution to the Ricci scalar will be dominated by that of the cosmological constant and we could test such a prediction.



\begin{acknowledgments}

I thank M. Sasaki for insightful discussions and financial support for the participation to the YITP Gravity and Cosmology 2012 workshop.

\end{acknowledgments}

\end{document}